*V. V. Isakevich,*[1] *L.V.Grunskaya,*[2] *D.V.Isakevich*[3]

# DETECTING SPECTRALLY LOCALIZED COMPONENTS OF LUNAR TIDE-FREQUENCY IN TIME-SERIES OF THE ELECTRIC FIELD VERTICAL COMPONENT OF THE EARTH ATMOSPHERE BOUNDARY LAYER[4] [5]

Using the signal eigenvectors and components analyser (Grunskaya L.V., Isakevich V.V., Isakevich D.V. the RF Utility Model Patent 116242 of 30.09.2011) made it possible to detect non-coherent complex-period components localized at lunar tide frequencies in the time-series of the electric field vertical component of the Earth atmosphere boundary layer. The detected components are unobservable by means of spectral analysis quadrature scheme; they are modulated with multiple frequencies of 1year$^{-1}$ and have RMS values 0.6–1.1 V/m. The probability of the detected effects being pseudo-estimates is not more than 2.5 · $10^{-4}$.

One of the tasks posed when exploring the electric field variations in the Earth atmosphere boundary layer was related to the study of how the electric field is affected by tide process. At the first stage the conventional quadrature scheme of spectrum evaluation [1] was used for detecting some periodic lunar tide frequency components in the time-series of the electric field vertical component of the Earth atmosphere boundary layer.

Long-term observation has made it clear that increasing the analysis time-range when using the conventional quadrature scheme does not lead to detecting the influence of lunar tides on the electromagnetic field of the Earth surface layer. This fact (as it was demonstrated later [2]) is a result of the sought components being non-coherent. For this reason the further research of the tides influence on the Earth electromagnetic field was carried out using the signal eigenvectors and components analyser [3] that proved its efficiency when detecting quasi-periodic energetically non-dominant components in experimental time-series.

This article presents the investigation results for four long-term observation time-series (TS) of the vertical component of the atmosphere boundary layer electric field: data from the testing ground of the VlSU Department of General and Applied Physics (2003–2009); data from geophysical observatories of Dusheti (1976–1980), Voyeikovo (1966–1995), and Verkhneye Dubrovo (1974–1995).

The TS were treated using the signal eigenvectors and components analyser (SEV&CA) [3], whose scheme is presented at figure 1. The analysed time-series ($1 \le t \le N$) is the input of scaling block 1. Block 2 calculates the covariance matrix $R$ of the time-series for the given analysis range of M counts of TS. Block 3 calculates eigenvectors and eigenvalues ($1 \le i \le M$). Block 4 performs the analysis of eigenvectors and eigenvalues in order to detect the characteristic signs of the sought physical phenomena and assessing their parameters. The covariance matrix $R$ in block 2 is calculated using the following formula:

$$R = \frac{YY'}{N-M+1} ,$$

Where $Y$ is the so-called trajectory matrix [4].

$$Y = \begin{bmatrix} X_1 & X_2 & \vdots & X_{N-M+1} \\ X_2 & X_3 & \vdots & X_{N-M+2} \\ X_3 & X_4 & \vdots & X_{N-M+3} \\ \ldots & \ldots & \ldots & \ldots \\ X_M & X_{M+1} & \vdots & X_N \end{bmatrix}.$$

The columns of the trajectory matrix $Y$ are the segments of TS got by moving along the latter the analysis range of M-counts length with a step =1.

---

1  E-mail: businesssoftservice@gmail.com
2  E-mail: grunsk@vlsu.ru
3  E-mail: voiceofhope@yandex.ru
4  The work is carried out with the financial support of the State Assignment 2014/13, 2871 and the RFRF Grant 14–07–97510/14.
5  The calculations were carried out using the free software `GNU Octave`, `CeCILL Scilab`, `GNU G95`.

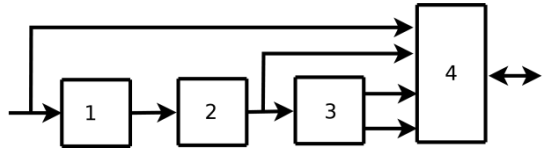

*Figure 1: The signal eigenvectors and components analyser [3].*
*1 – scaling block, 2 – covariance matrix calculating block. 3 – eigenvectors and eigenvalues calculating block. 4 – TS signs and components analyser.*

TS for Every M-length analysis range can be presented in eigenvector basis of the covariance matrix R (forming the orthonormal basis [5]) defined by the equation.

$$R\vec{\psi}_i = \lambda_i \vec{\psi}_i, 1 \leqslant i \leqslant M.$$

It's easy to demonstrate that the energy mean value *E* for the TS observed in the analysis range is defined by the equation.

$$E = \sum_{i=1}^{M} r_{i,i} = \text{Tr}R = \sum_{i=1}^{M} \lambda_i,$$

Where $r_{i,i}$ —the *i*-th diagonal element of the covariance matrix *R*,
*TrR* — the trace of the covariance matrix *R*.

Thus, eigenvectors are orthonormal non-correlated components of TS analysed in the M-length range. Their relative contribution to the mean energy *E* of the analysed TS is

$$\lambda_i^{\text{норм.}} = \frac{\lambda_i}{\sum_{i=1}^{M} \lambda_i}.$$

The series of eigenvalues (sorted descending) we'll call the eigenvalues spectrum (EVS) and the corresponding series (5) — the normalized eigenvalues spectrum (NEVS).

SEV&CA belongs to adaptive basis analysers' class. It means that the basis in which TS is represented for the finite analysis range depends on the TS itself. It's easy to demonstrate (using well-known linear algebra theorems [5]) that (for the given accuracy of representation) representing TS in eigenvectors (EV) basis takes the least number of components, i.e. it has the maximum expressiveness. Furthermore, EV themselves contain the information of the analysed TS structure. Thus, if TS has spectrally localized components, its EVs will also contain those components which can be detected by conventional means of spectral analysis. It has been demonstrated [6] that the EV spectral analysis has higher sensitivity than the conventional TS spectral analysis, especially when the components are not absolutely coherent. SEV&CA often makes it possible to define the type of the analysed TS according to the form of its NEVS [3].

Using SEV&CA is expedient when it's difficult to apply the conventional quadrature scheme of spectral analysis, e.g. because of the TS non-coherence. Figures 2 and 3 demonstrate in double-log scale the amplitude estimates of the spectrum component having the solar tide frequency (figure 2) and the lunar tide frequency (figure 3). The amplitude estimates have been got using the conventional quadrature scheme for TS from the Voyeikovo station.

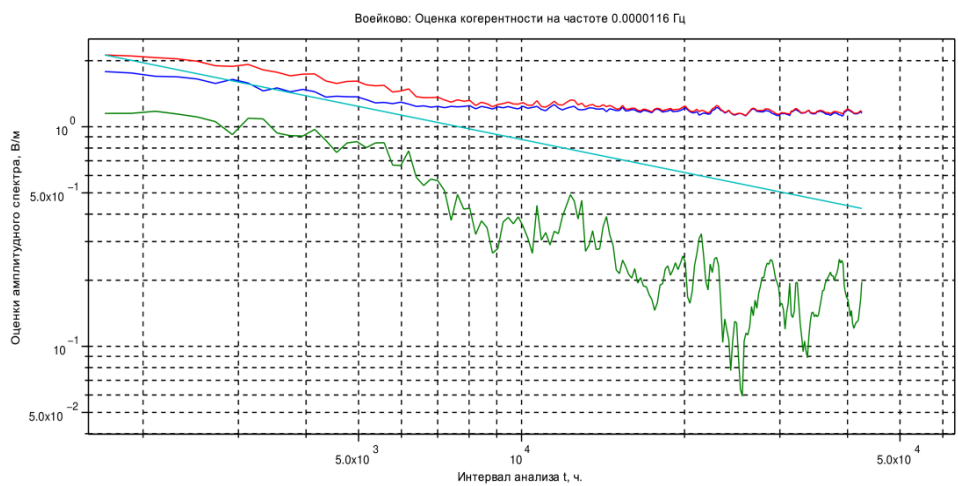

*Figure 2: The estimate of the spectrum component of solar tide frequency vs.analysis time-range when using quadrature scheme. An upper curve is the RMS value, a middle curveo is sample mean, a lower curve is sample variance. The straight line* $\sim 1/\sqrt{t}$ *where t is the time-range length.*

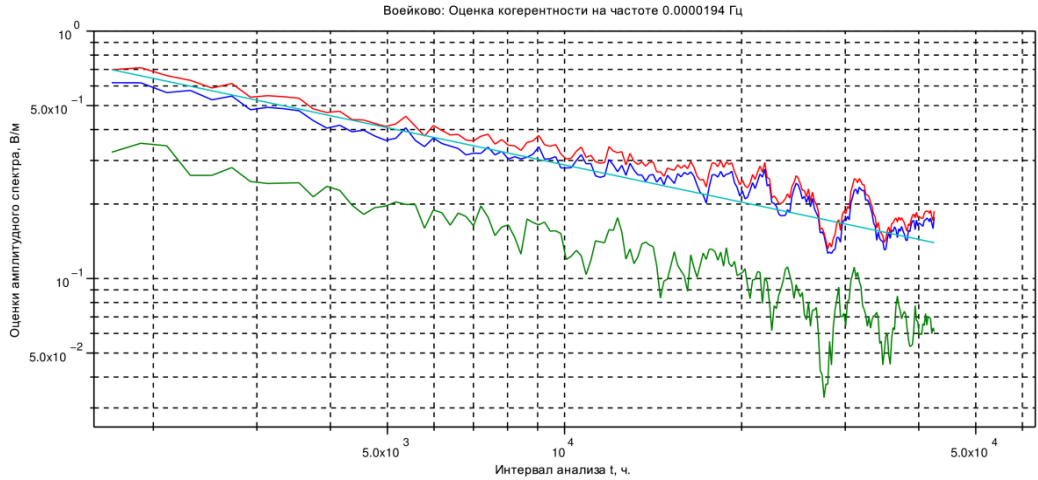

*Figure 3: The estimate of the spectrum component of lunar tide frequency vs. analysis time-range when using quadrature scheme. An upper curve is the RMS value, a middle curve is sample mean, a lower curve is sample variance. The straight line ~ $1/\sqrt{t}$ where t is the time-range length.*

As it can be seen at figure 2, using quadrature scheme presents no problems when detecting the component of the solar tide frequency — the sample mean and the RMS values converge to some fixed nonzero value and the spectral estimate variance decreases monotonically when increasing the analysis range. Quite the opposite situation is observed for the component of lunar tide frequency (figure 3). While the analysis time-range $t$ is increasing the sample mean and the RMS values are decreasing monotonically. It means that there's no coherent component localized at the lunar tide frequency. The relations showed at figures 2 and 3 are typical for all the solar tides and lunar tides at every observation point.

In order to detect non-coherent components localized at lunar tide frequencies block 2 of SEV&CA used covariance matrixes 1000·1000 built on the basis of the analysed TS trajectory matrixes. In block 4 of SEV&CA the eigenvectors were treated by spectral analysis using the standard FFT procedure.[6] For every EV the coherence index (CI) was calculated as a ratio of FFT amplitude at the lunar tide frequency to the FFT amplitude mean value (similar to "signal-noise" ratio). Using FFT for spectral analysis limits the CI value to $M/2$ (for this work is 500). All the eigenvectors having the amplitude spectrum maximum at the lunar tide frequency were detected. The CI for detected EV are presented in the last column of table 1.

Table 1: RMS values of non-coherent uncorrelated components of lunar tide frequencies calculated using SEV&CA
($M$=1000, $\Delta t$=1 hr.)

| Tide | Period, hr. | Station | EV number | RMS value, V/m | CI |
|---|---|---|---|---|---|
| 4* | 4*14.3261 | Voyeikovo | 134 | 0.74 | 77.89 |
| | | Verkhneye Dubrovo | 130 | 0.63 | 16.33 |
| | | Dusheti | 107 | 0.65 | 31.59 |
| | | VlSU | 91 | 0.66 | 117 |
| 4* | 4*23.0646 | Voyeikovo | 89 | 0.97 | 107 |
| | | Verkhneye Dubrovo | 82 | 0.89 | 113 |
| | | Dusheti | 119 | 0.63 | 57 |
| | | VlSU | 61 | 1.02 | 157 |
| 4* | 4*24.8724 | Voyeikovo | 80 | 1.03 | 147 |
| | | Verkhneye Dubrovo | 77 | 0.92 | 134 |
| | | Dusheti | 124 | 0.62 | 31 |
| | | VlSU | 47 | 1.10 | 264 |
| 4* | 4*12.65 | Voyeikovo | 167 | 0.65 | 67 |
| | | Verkhneye Dubrovo | 135 | 0.62 | 59 |
| | | Dusheti | 131 | 0.61 | 58 |
| | | VlSU | 214 | 0.28 | 25.5 |
| 4*, | 4*25.8176, 25.71(6) | Voyeikovo | 62 | 1.13 | 132 |
| | | Verkhneye Dubrovo | 85 | 0.88 | 109 |
| | | Dusheti | 40 | 0.79 | 9 |
| | | VlSU | 49 | 1.1 | 190 |
| 3* | 3*18.9891 | Voyeikovo | 112 | 0.83 | 15 |
| | | Verkhneye Dubrovo | 101 | 0.77 | 32 |
| | | Dusheti | 134 | 0.61 | 28 |
| | | VlSU | 179 | 0.35 | 52.3 |

Table 1 contains data about EVs having the spectrum maximum at lunar tide frequencies. For every tide and station there are data concerning the only EV, the one having the maximum CI. For every such EV the RMS value of its principle component was calculated as a product of TS RMS value by . This method is given in details in [2]. As it can be seen in table 1 the RMS values of components spectrally localized at lunar tide frequencies are located in the range of 0.6–1.1 V/m and are similar for TS obtained at different stations. Table 1 contains only those lunar tides whose frequencies differ significantly from the "gravitation beacons" frequencies. The influence of the latter on the electromagnetic field will be analysed in the next article.

The examples of eigenvectors and their normalized amplitude spectra for some lunar tides and observation points are shown at figures 4, 5 and 6. It can be seen that the normalized amplitude spectra are localized near lunar tide frequencies.

In [7] it is demonstrated that if the SEV&CA input receives a polyharmonic TS then the EV of its covariance matrix can have only harmonic components of the frequencies contained in the initial polyharmonic TS. Though the analysed TS are not polyharmonic, it should be expected that their EV and amplitude spectra localized at lunar tide frequencies also contain some components localized at other frequencies (including frequencies of other lunar tides). Let's consider the results shown at figures 4, 5, 6 from this viewpoint.

Figure 4 shows the EV and its amplitude spectrum having maximum at the frequency $M_1$. As it was mentioned earlier the frequency discrete $f$=0.278 mcHz; the observed spectrum width is not less than 4$f$, it includes tide frequencies , and , positioned symmetrically with respect to the tide frequency (see fig. 8). In our opinion this fact is the reason of the EV regular shape (characteristic for beats with close frequencies) as well as the regular shape of the EV amplitude spectrum. At figure 5 there is a peak to the left of the EV amplitude spectrum main maximum. The peak relates to the lunar tide frequency (accurate to $t$).

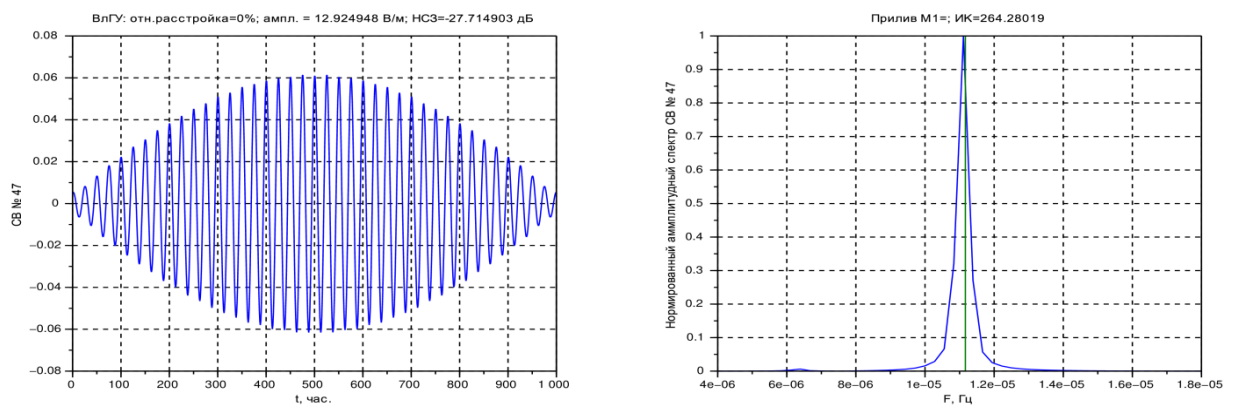

*Figure 4: The eigenvector localized at lunar tide frequency and its normalized amplitude spectrum. VlSU testing ground. Continuous vertical line relates to the lunar tide frequency.*

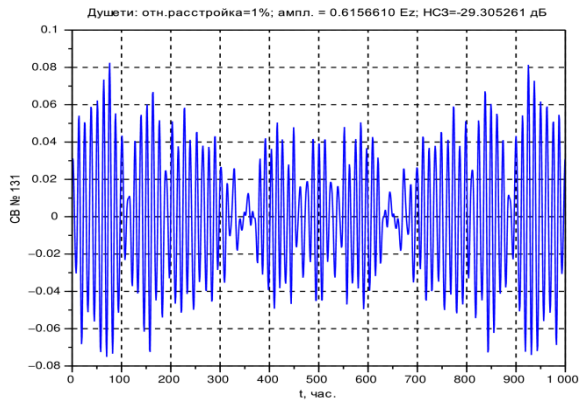
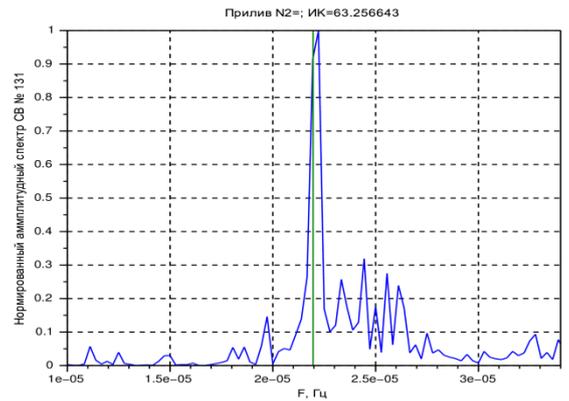

*Figure 5: The eigenvector localized at lunar tide frequency and its normalized amplitude spectrum. Dusheti. Continuous vertical line relates to the lunar tide frequency.*

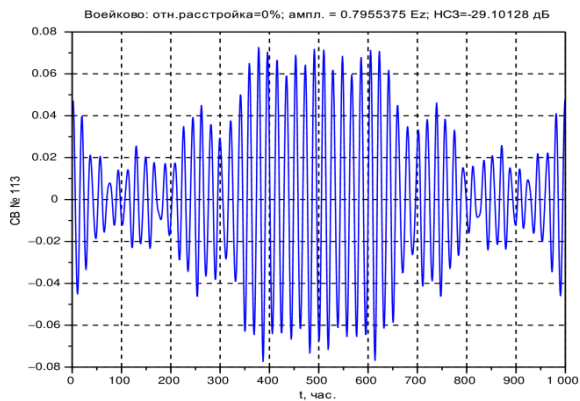
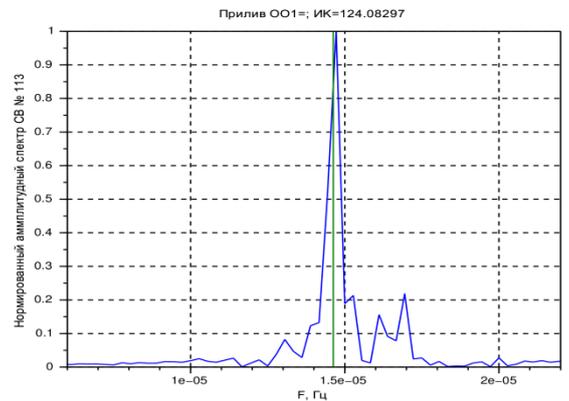

*Figure 6: The eigenvector localized at lunar tide frequency and its normalized amplitude spectrum. Voyeikovo. Continuous vertical line relates to the lunar tide frequency.*

It was demonstrated that the amplitudes of components localized at lunar tide frequencies change periodically with frequencies that are multiples of 1year$^{-1}$. For that purpose the TS normalized projections to the EV spectrally localized at lunar tide frequencies were compared to the test polyharmonic series having components of unit amplitudes and frequencies multiple to 1year$^{-1}$. Figure 7 shows the example of amplitude spectrum for the normalized projection of TS (Dusheti station) to the EVs localized at lunar tide frequency.

Detecting of amplitude modulation confirms that the annual changing of the Earth position relative to the Sun affects the TS components localized at lunar tide frequencies. It should be mentioned that modulations of that kind are characteristic for most EVs of TS. It indicates that those modulations are generated by physical phenomena also subjected to annual cycles.

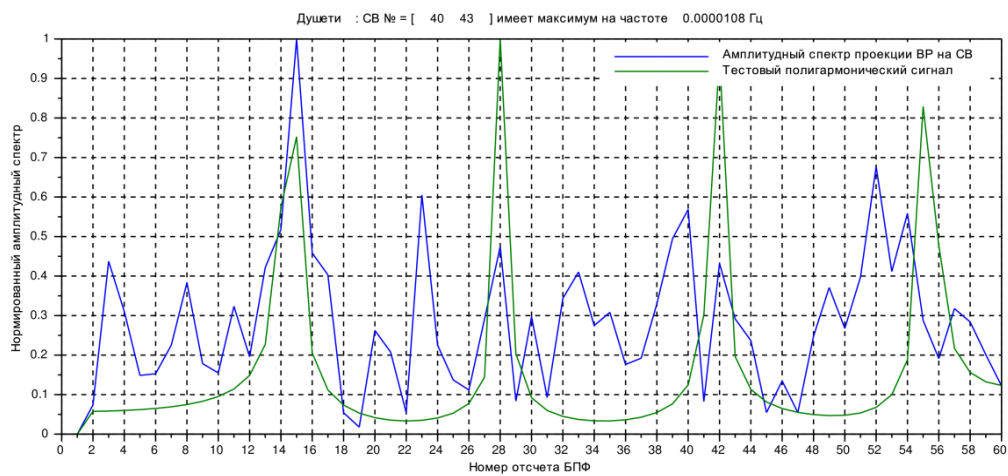

*Figure 7: The amplitude spectrum of the normalized projection of TS (Dusheti station) on the subspace made by EVs 40 and 43 spectrally localized at lunar tide frequency, in comparison to the amplitude spectrum of the test polyharmonic TS with component periods 1, 1/2, 1/3, 1/4 years.*

The question that comes naturally: can eigenvectors localized at lunar tide frequencies arise by chance? If this hypothesis is confirmed then all the previously described results can be characterized as pseudo-estimates. Thus, confirming or refuting the chance hypothesis becomes the problem of particular importance. We use standard approach [8] in order to estimate the "false alarm" probability (i.e. the probability of the chance hypothesis being true). As we know, this probability is defined, on the one hand, by the decision rule and on the other hand — by statistical properties of values used in the rule. As a decision rule we can use the most simple sampling rule "*k of K*" stated further.

**Rule 1**. *Let it be K values of CI ($1 \leq j \leq K$) got for EVs selected during the analysis of all accessible TS for all the studied lunar tide frequencies. If of these values exceed their corresponding thresholds then the chance hypothesis is refuted.*

Using rule**Ошибка! Источник ссылки не найден.** the chance hypothesis can be formulated as follows:

**Hypothesis 1.** *The observed exceeding of the thresholds by of K coherence indexes has occurred by chance.*

The alternative hypothesis is the following:

**Hypothesis 2.** *The observed exceeding of the thresholds by of K coherence indexes has not been caused by chance.*

As thresholds we take quantiles corresponding to the *p* probability of their exceeding by the coherence indexes. Then the probability of hypothesis being true (the "false alarm" probability, $P_{лт}$) under the binomial distribution can be estimated by the following relation:

$$P_{лт}(k_0, K, h) = P\{k \geqslant k_0 | \text{Гипотеза 1}\} = \sum_{k=k_0}^{K} C_K^k p^k (1-p)^{K-k},$$

When are equal to CI median (i.e. *p*=1/2) and the relation (6) is reduced to inequality

$$P_{лт}(k_0, K) \leqslant \frac{C_K^k}{2^K}(K - k_0 + 1).$$

This inequality has a corresponding simple decision rule.

**Rule 2.** *If we have observed not less than (of K possible) CI values that exceed the CI median value then the hypothesis alternative to 1 should be accepted.*

In order to use rule and to estimate by relation (7) the median values of CI at lunar tide frequencies should be estimated. For this purpose two calculating experiments were carried out.

In the first experiment for each of the four initial TS we built an ensemble of TS segments having the length of counts (ensemble size 50–60). For every segment a covariance matrix was built, EVs and their FFTs were calculated. For every FFT we found the position of the amplitude spectrum maximum and the CI. Then for every position of the amplitude spectrum maximum a CI sample was built and its median was estimated. Thus we've got the relation between CI median and the frequency of EVs amplitude spectrum maximum. This relation, typical for all the TS, is shown at figure 8.

Similarly, in the second calculating experiment we've found the relation between the CI median and the frequency of EVs amplitude spectrum maximum of the TS segments ensemble for white Gaussian noise having the length of counts (ensemble size = 200). This relation is also shown at figure 8. It's clearly seen that CI median values for WGN are significantly less than that for TS.

Using relation (7), data about the CI of EVs localized at lunar tide frequencies as well as median values estimates got in the first calculating experiment makes it possible to get the estimate for , *K*=45 as the inequality $P_{лт}(36, 45) \leqslant 2.5 \cdot 10^{-4}$.

The second calculating experiment does not directly concern the analysed TS. But it gives the answer to the frequently heard statement of our critics: «If you consider Gaussian noise, your SEV&CA will give the same result as you've got analyzing TS». The calculating experiment demonstrates that analyzing Gaussian noise (WGN) by SEV&CA gives the probability of CI exceeding threshold $P_{лт}(43, 45) \leqslant 8.4 \cdot 10^{-11}$.

So, the analysis of long-term time series at four stations (Voyeikovo, Dusheti, Verkhnyaia Dubrova, VlSU testing ground) carried out by means of signal eigenvectors and components analyser makes it possible to assert that non-coherent complex-periodic components were detected at lunar tide frequencies, the "false alarm" probability being not more than .

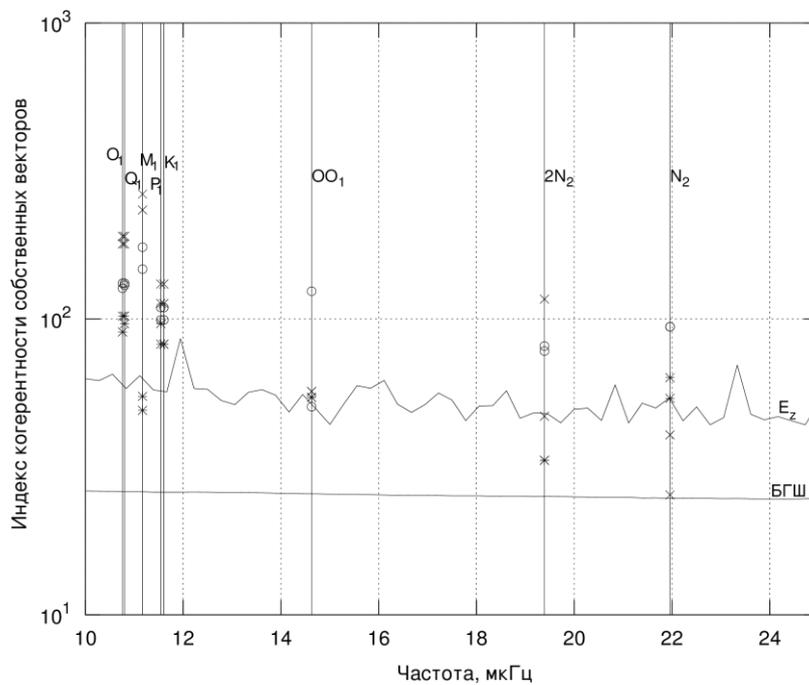

*Figure 8: CI vs. position of EV amplitude spectrum maximum. The result of calculating experiment for TS segments (upper curve) and for Gaussian white noise (WGN) segments of the same length (lower curve) in comparison to CI of TS eigenvectors spectrally localized at lunar tide frequencies. Vertical lines show lunar tide frequencies*
*x — VlSU; \* — Dusheti; o — Voyeikovo.*

The results were obtained using the authors' registered software [9,10].

**Conclusions.**
1. It is demonstrated that those components of the time-series (TS) of electric field vertical component () in the Earth boundary layer which are localized at lunar tide frequencies are non-coherent, unlike those at solar tide frequencies. This is the reason why increasing analysis time-range leads to decreasing of estimates got by means of conventional quadrature scheme. So, quadrature scheme is not applicable for detecting components localized at lunar tide frequencies.
2. Unlike the conventional quadrature scheme, the signal eigenvectors and components analyser [3] (SEV&CA) makes it possible to detect non-coherent components localized at lunar tide frequencies. SEV&CA uses the TS representation in the orthonormal adaptive basis of the TS covariance matrix eigenvectors (EV). Eigenvectors are fully defined by their covariance matrix and hence contain the information about TS internal links in the given analysis time-range. In the work presented the length of the time-range is TS counts with discretization time of 1 hr. Studying the TS proved that EVs have complex-periodic structure. The EVs spectral analysis makes it possible to detect EVs localized at the frequencies characteristic for the TS.
3. Every EV has its corresponding eigenvalue (EVal) which is equal to the energy of the EV-related component averaged in the analysis range. Eigenvalues , taken relative to their sum and arranged decreasingly form the normalized EVal spectrum (NEVS). Eigenvectors then are numbered according to their decreasing EVal. Values in NEVS define the energy relative contribution of the TS uncorrelated componentscorresponding to EV. The component RMS value is then defined as a product of TS RMS value by It is demonstrated that components localized at lunar tide frequencies have RMS values in the range 0.6–1.1 V/m.
4. The degree of EV amplitude spectrum localization near lunar tide frequencies was assessed by coherence index (CI) whose value is equal to the FFT amplitude at the tide frequency related to its mean value. CI can be considered similar to "signal-noise" ratio, its value is limited by *M*/2 (in this work — 500). For the studied TS and considered lunar tide frequencies the CI experimental estimates are in the range 9–264.
5. The results of the calculating experiment using TS made it possible to build the relation of CI median vs. frequency of EV amplitude spectrum maximum. For those amplitude spectrum maximums that relate to lunar tide frequencies the CI median is in the interval 47–60. In the similar way another calculating experiment defined the relation of CI median vs. frequency of EV amplitude spectrum maximum for the sample of segments of Gaussian "white" noise TS (WGN). It should be mentioned that CI median values for WGN are less than those for the segments of the studied TS.
6. Comparing the particular values of CI for TS and considered lunar tide frequencies to the median values of CI makes it possible to estimate using binomial distribution the "false alarm" probability (i.e. the probability of CI particular value exceeding CI median value by chance). For lunar tide frequencies 36 of 43 CI particular values exceed median. This corresponds to the "false alarm" probability not more than $2.5 \cdot 10^{-4}$
7. It is demonstrated that the amplitudes of components localized at lunar tide frequencies, as well as most of the other non-correlated components, change periodically having frequencies that are multiples of 1 year$^{-1}$.

Thus, using SEV&CA confirms with high reliability the presence of non-coherent complex-periodic components spectrally localized at lunar tide frequencies in the Earth electric field vertical component. Using SEV&CA also makes it possible to estimate their coherence and RMS values.

Isakevich Valery Victorovich, Doctor of Philosophy, Senior Researcher, Department of General and Applied Physics, Vladimir State University, ul. Gorkogo, 87, Vladimir, 600000, Russia; Chief Development Officer, BusinessSoftService Ltd, ul. B. Moskovskaya, 61, Vladimir, 600001, Russia.
E-mail: businesssoftservice@gmail.com
Grunskaya Lyubov Valentinovna, Doctor of Science, Professor, Department of General and Applied Physics, Vladimir State University, ul. Gorkogo, 87, Vladimir, 600000, Russia.
E-mail: grunsk@vlsu.ru
Isakevich Daniel Valerievich, Engineering Researcher, Department of General and Applied Physics, Vladimir State University, ul. Gorkogo, 87, Vladimir, 600000, Russia; Chief Technical Officer, BusinessSoftService Ltd, ul. B. Moskovskaya, 61, Vladimir, 600001, Russia.
E-mail: voiceofhope@yandex.ru.